\documentclass[9pt,twocolumn,twoside,lineno]{pnas-new}

\templatetype{pnasresearcharticle} 

\title{Learning to predict target location with turbulent odor plumes}

\author[a,b,c]{Nicola Rigolli}
\author[b,c]{Nicodemo Magnoli}
\author[c]{Lorenzo Rosasco}
\author[a]{Agnese Seminara}

\affil[a]{Universit\'e C\^ote d'Azur, CNRS, Institut de Physique de Nice, UMR7010, Parc Valrose 06108, Nice, France}
\affil[b]{Department of Physics, University of Genova, via Dodecaneso 44, 16144 Genova, Italy}
\affil[c]{INFN, sezione di Genova, Via Dodecaneso 33, 16146 Genova, Italy}
\affil[c]{MaLGa, DIBRIS, University of Genova, via Dodecaneso 43, 16144 Genova, Italy}

\leadauthor{Rigolli} 

\significancestatement{Organisms solve sophisticated olfactory problems even when odor cues are broken by turbulence and chemical gradients become useless.
Here we analyze turbulent dynamics to explore olfactory predictions, i.e.~how can an agent predict the position of a target using a short extract of the odor it releases.
We find that measures of timing and intensity of odor detections enable olfactory predictions. But even within a well defined fluid flow, the quality of predictions varies in space. Variations occur because the effect of turbulence becomes increasingly severe as the odor travels away from its source. Our results demonstrate that to maintain predictive power, animals must rely on different measures as they move which -we predict- will be reflected in both behavior and neural representations of odor dynamics. 
}

\authorcontributions{Please provide details of author contributions here.}
\authordeclaration{The authors declare no conflict of interest.}
\correspondingauthor{\textsuperscript{2}To whom correspondence should be addressed. {E-mail: agnese.seminara@ univ-cotedazur.fr}}

\keywords{Olfaction $|$ Turbulence $|$ Prediction $|$} 

\begin{abstract}
Animal behavior and neural recordings show that the brain is able to measure both the intensity of an odor and the timing of odor encounters. However, whether intensity or timing of odor detections is more informative for olfactory-driven behavior is not understood. 
To tackle this question, we consider the problem of locating a target using the odor it releases.
We ask whether the position of a target is best predicted by measures of timing \emph{vs} intensity of its odor, sampled for a short period of time. 
To answer this question, we feed data from accurate numerical simulations of odor transport to machine learning algorithms that learn how to connect odor to target location. 
We find that both intensity and timing can separately predict target location even from a distance of several meters; however their efficacy varies 
with the dilution of the odor in space.
Thus organisms that use olfaction from different ranges may have to switch among different modalities. This has implications on how the brain should represent odors as the target is approached.
We demonstrate simple strategies to improve accuracy and robustness of the prediction by modifying odor sampling and appropriately combining distinct measures together. 
To test the predictions, animal behavior and odor representation should be monitored as the animal moves relative to the target, or in virtual conditions that mimic concentrated \emph{vs} dilute environments.
\end{abstract}

\dates{This manuscript was compiled on \today}
\doi{\url{www.pnas.org/cgi/doi/10.1073/pnas.XXXXXXXXXX}}

\begin{document}

\maketitle
\thispagestyle{firststyle}
\ifthenelse{\boolean{shortarticle}}{\ifthenelse{\boolean{singlecolumn}}{\abscontentformatted}{\abscontent}}{}


Most macroscopic organisms detect odors in intermittent bursts, that may be separated by extended regions with no odor. Organisms leverage this complex dynamics efficiently for diverse tasks, including locating and identifying an odor source \cite{Murlisetal1992,mafraneto_Carde1994,Vickers2000,Riffeletal2014,Acheetal2016,schaefer2021}. However, what are the most informative features of intermittent odor cues remains largely unclear. There are two broad classes of measures that quantify the dynamics of olfactory cues: those that depend on odor intensity including e.g.~odor gradients in space or time, and those that do not depend on odor intensity but only on its timing, i.e.~on whether the odor is on or off regardless of its concentration. To compute quantities that depend on odor intensity, an accurate representation of the odor is needed. In contrast, measuring the timing of odor detection simply requires to mark at all times whether the odor is on or off, thus a binary switch is sufficient. 
\\
Behavioral evidence suggests that animals use both intensity and timing of odor encounters for olfactory navigation~\cite{Bakeretal2018}. 
At close range, mammals appear to compare odor intensity either across nostrils or across sniffs~\cite{Catania2013,Gireetal2016,Findleyetal2020}. 
On the other hand, mounting evidence suggests timing of odor detection also plays a key role for olfactory navigation~\cite{Acheetal2016}: moths respond to odor pulsed at specific frequencies~\cite{Riffeletal2014,Vickersetal2001}; fruit flies respond to timing since last odor detection~\cite{vanBreugel_Dickinson2014,Demiretal2020}; lobsters and sharks compare odor arrival time across their paired olfactory organs and orient toward the side that detected the odor first~\cite{BasilAtema1994,GardinerAtema2010}; many organisms will move upwind upon detection of an odor~\cite{Kennedy_Marsh1974,Murlisetal1992,Stecketal2012}. \\
Neural recordings upon stimulation with intermittent odor cues 
confirm that the brain of many animals is able to record information both about intensity (and its derivatives) as well as timing of odor encounters (most information comes from work on arthropods~\cite{NagelWilson2011,Vickersetal2001,Brownetal2005,Gorur-Shandilya2017,Jacobetal2017,Riffeletal2014}, but see also~\cite{Parabuckietal2019,Lewisetal2021}).  
For example, when insects are presented with intermittent odor cues, information about intensity and timing is recorded in their antennal lobe~(see e.g.~\cite{Vickersetal2001,Brownetal2005}). 
Odors that mimic natural intermittency elicit a response that preserves an accurate measure of timing in fruit flies and moths~\cite{Gorur-Shandilya2017,Jacobetal2017}. 
In lobsters, bursting olfactory neurons encode specifically for the time between successive odor encounters, see~\cite{Parketal2014,Parketal2016} and references therein.
Interestingly, the neural activity varies considerably with the dynamics of the odor cues~\cite{NagelWilson2011,Vickersetal2001,Lewisetal2021}, but how intermittency of an odor affects its neural representation is not well understood. 
\\

This evidence suggests animals are able to identify when they detect an odor as well as how intense it is; but whether they record and rely on both kinds of information is not understood. From a physical perspective, these two measures clearly provide information about source location. Indeed, we know from theoretical~\cite{ss00,fgv01,prx} and experimental~\cite{Murlisetal1992,MOORE2004} work that turbulence causes the odor to be distributed in highly intermittent patches separated by blanks with no odor. Both intensity and timing of these intermittent bursts vary depending on the location of the source~\cite{prx}, as early recognized by~\cite{Atema1996}, thus can be used to infer source location or navigate to it~\cite{infotaxis,Schmukeretal2016,boieetal2018,Leathers2020,michaelisetal2020}.

\begin{figure*}[h!!]
\begin{center}
\includegraphics[width=0.75\textwidth]{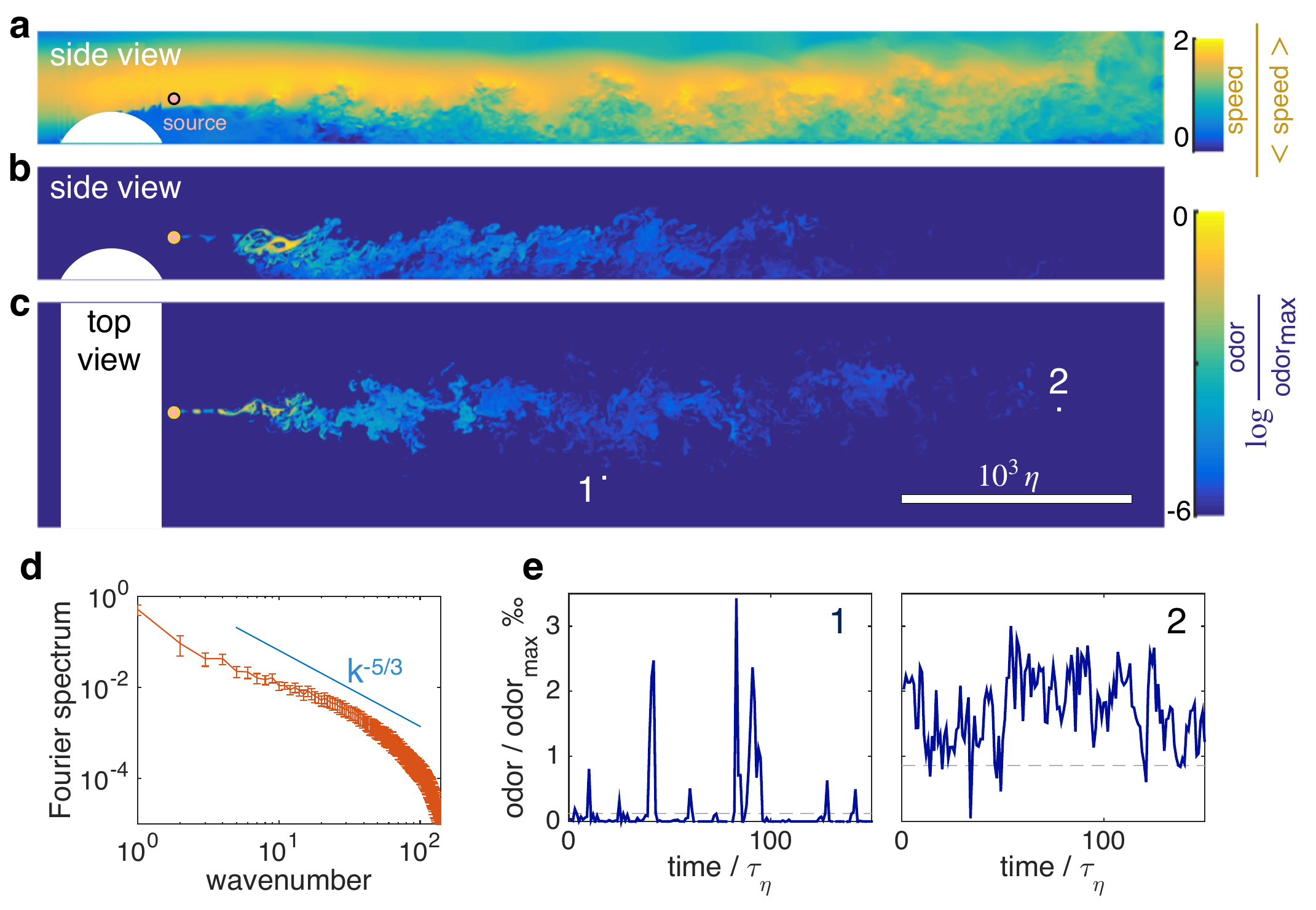}
\caption{Turbulent odor cues are patchy and intermittent. 
Snapshot of streamwise velocity (a) in a vertical plain at mid channel; odor snapshot side view at mid channel (b) and top view at source height (c). White regions mark the cylindrical obstacle. Snapshots are obtained from direct numerical simulations of the Navier-Stokes equations and the equation for odor transport (see Materials and Methods and parameters summarized in Table~\ref{tab:par}). (d) Spectra of odor fluctuations compared to the prediction for turbulent signals. (d) Typical time courses of the odor cues at locations labeled with $1$ and $2$ in c, visualizing noise and sparsity, particularly at location 1.}
\label{fig:snapshots}
\end{center}
\end{figure*}

Here we ask what salient features of turbulent odor signals best predict the location of the odor source and specifically compare quantities related to intensity \emph{vs} timing of odor encounters. We first compose a dataset of realistic odor fields at scales of several meters using accurate state-of-the-art fluid dynamics simulations. We then develop machine learning algorithms that predict source location based on these synthetic odor fields. 
We find that measures of odor temporal dynamics based on a short memory span (down to about 1 second) hold information about source location. Close to the source or close to the substrate, measures of intensity predict distance better than measures of timing; but this ranking is reversed at further distance from the source or from the substrate. Pairing the two kinds of measure improves dramatically the quality of the prediction robustly across all datasets, whereas pairing two measures of intensity or two measures of timing is either useless or detrimental. 

Our results demonstrate that timing and intensity are complementary attributes of odor dynamics and are most effective in more dilute and concentrated conditions respectively. 
These different conditions exist in different portions of space because odor gets transported, mixed and diluted by the fluid. 
As a result, the spatial range of operation of a living organism constrains the solutions it may evolve to 
make predictions with turbulent odors.

\section*{Results}

Odor cues at several meters from the source are often turbulent. Figure~\ref{fig:snapshots}a-c and show snapshots of the velocity field and odor cues in space, resulting from direct numerical simulations of the turbulent flow in a channel of length L, width W and height H (also see Supplementary Movie). Air flows from left to right at a mean speed $U_b$ and hits a cylindrical obstacle that generates turbulence. The height of the obstacle is $H/4$ and tunes the intensity of turbulent fluctuations relative to the mean velocity. The odor field is emitted from a concentrated source downstream from the obstacle; it develops as a meandering filament that fluctuates as it travels downstream and soon breaks into discrete pockets of odor (whiffs) separated by odor-less stretches (blanks) (Figure~\ref{fig:snapshots}b-c,e). The spectrum of odor fluctuations is consistent with $k^{-5/3}$ scaling typical of turbulent transport (Figure~\ref{fig:snapshots}d), which is confirmed by the sparsity of odor cues in time (Figure~\ref{fig:snapshots}e). Note that depending on the sampling location, odor may be more or less sparse (compare for example Figure~\ref{fig:snapshots}e left and right). All parameters and methods are summarized in Table~\ref{tab:par} and Materials and Methods.
\\

Do odor cues bare information about source location meters away from the source? To answer this question, we develop supervised machine learning algorithms that learn the relationship between the input (odor) and the distance from the source (output) from a large dataset of examples. 
In order to dissect what are the best predictors of source location and how ranking depends on the statistics of the odor, we need to detail more specifically the input and output of the algorithm.\\

\begin{figure}[h!!]
\begin{center}
\includegraphics[width=0.5\textwidth]{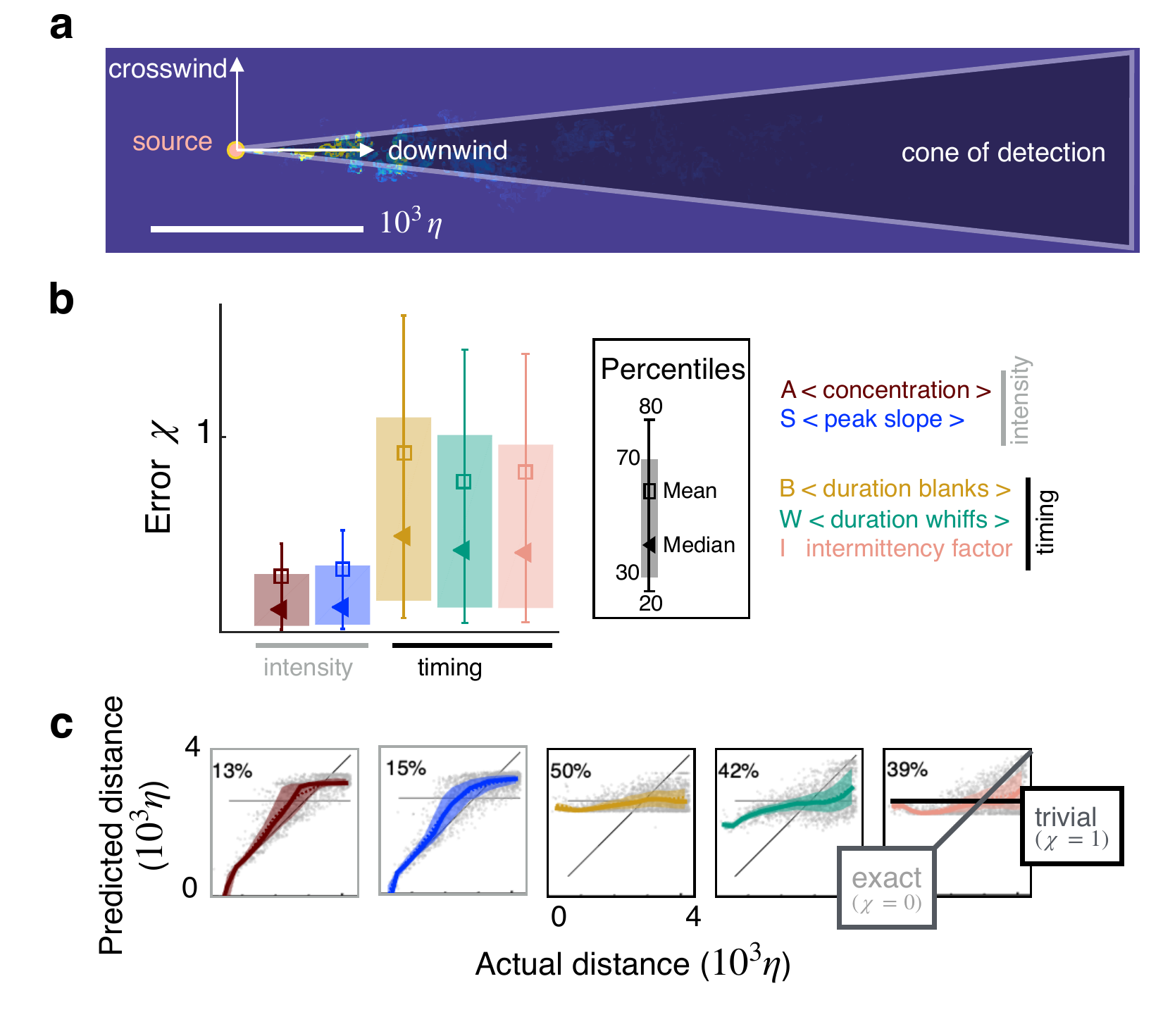}
\caption{Individual features enable inference in two dimensions. (a) Sketch of the geometry. (b) Test error $\chi$ for inference using individual features as input. (c) Predicted \emph{vs} actual distance for inference. 
Prediction for representative test points (grey circles); 30$^{th}$ to 70$^{th}$ percentile (patch, same color code as in (b)); trivial prediction $f(\mathbf{x})=<y>_{\text{test}}$ (solid horizontal line, corresponds to $\chi=1$); exact prediction (bisector, corresponds to $\chi=0$); dispersion away from the bisector visualizes the prediction error. 
Results are obtained with a supervised learning algorithm based on regularized empirical risk minimization (Materials and Methods). Each input datum $x_i$ is one individual scalar feature computed from the time course of odor concentration measured at location $\mathbf{z}_i$ at 100 evenly spaced time points with sampling frequency $\omega=1/\tau_\eta$, where $\tau_\eta$ is Kolmogorov time. The training/test set are composed of $N=5000$ and $N_t=13500$ data points respectively. 
}
\label{fig:cone}
\end{center}
\end{figure}
To design the input we start with the odor concentration field $c(\mathbf{z},t)$ which varies stochastically in space and time as a result of turbulent transport. 
Here $\mathbf{z}=(z_1,z_2,z_3)$ is a location in the three dimensional space and $t$ is time. 
We focus on a plane at a fixed height, and consider the conical region where odor can be detected, the ``cone of detection'' (Figure~\ref{fig:cone}a). 
We first compose time series of the odor field; each time series is indicated with $\mathbf{c}_i$ and consists of the odor sampled at $M$ equally spaced times with frequency $\omega$ at a discrete location $\mathbf{z}_i$ within the cone of detection. 
Thus each time series is a vector $\mathbf{c}_i=(c(\mathbf{z}_i,t_i),...,c(\mathbf{z}_i,t_{i+M}))$, where $t_{i+M}-t_i=M/\omega$ is the temporal span of the time series, or memory.
From each time series $\mathbf{c}_i$ we calculate five features $x_i^1,...,x_i^5$, where $x_i^1$ is the temporal average of the concentration  during whiffs in the time series $\mathbf{c}_i$; $x_i^2$ is its average slope (time derivative of odor upon detection, averaged across whiffs within $\mathbf{c}_i$); $x_i^3$ is the average duration of blanks (stretches of time when odor is below detection within $\mathbf{c}_i$); $x_i^4$ is the average duration of whiffs (stretches of time when odor is above threshold within $\mathbf{c}_i$); and $x_i^5$ is the intermittency factor (the fraction of time the time series $\mathbf{c}_i$ is above threshold). The detection threshold is defined adaptively as discussed in Materials and Methods.   
Features $x^1$ and $x^2$ depend explicitly on odor concentration, whereas features $x^3$, $x^4$ and $x^5$ only depend on when the odor is on or off, but not on its intensity. To remark this difference, we refer to  $x^1$ and $x^2$ as intensity features, and $ x^3$, $x^4$ and $x^5$ as timing features. 
Our input $\mathbf{x}_i=(x_i^1,...,x_i^d)$ is composed of d-dimensional vectors of features and we will focus on $d=1,2,5$. 
We seek to infer distance from the source, thus our output $y$ is the coordinate of the sampling point $\mathbf{z}$ in the downwind direction, i.e.~$y=z_1$, with the source placed at the origin (see sketch in Figure~\ref{fig:cone}a). We refer to the supplementary material for results in the crosswind direction, $y=z_2$. 
We train the algorithm by providing $N$ examples of input-output pairs $(\mathbf{x}_i,y_i)$ selected randomly from the full simulation, and obtain the function that connects input and output: $y \approx f(\mathbf{x})$. \\

We propose a machine learning approach where  the different odor  features are ranked based on  their predictive power, rather than  their  fitting properties. Different data-sets of  odor/distance pairs are defined. The data-sets  differ in the way odor measurements are represented in terms of feature vectors. For each data-set we learn  a function to predict  the distance to target  given the corresponding odor  features. The predictive power of each function, and corresponding set of features, is then assessed. More precisely each  data-set is split in  a training  and a test set, as  custom in machine learning.  Training sets are used  to learn functions connecting odor to target location, whereas  test sets are used to assess their prediction properties. The training/test split is crucial since  the goal is to make good predictions on new, unseen  points, that are not within the training sets. From a modeling perspective, a flexible nonlinear/nonparametric approach based on kernel methods is contrasted and shown to be superior to  a simpler linear model (Supplementary Figure~S1).  A careful protocol based on hold-out cross-validation  is used to select the hyper-parameters of the considered learning models. We refer to  Materials and Methods for all details.\\ 

To illustrate the results we pick the two dimensional plane at height $H/4$ that contains the source. 
The first result is that individual features ($d=1$) bare useful information for two-dimensional source localization even at several meters from the source. Performance is quantified by the normalized squared error averaged over the $N_t$ points in the test set $\chi =  \sum_{i=1}^{N_t}[y_i-f(\mathbf{x}_i)]^2 /  \sum_{i=1}^{N_t}[y_i - \bar{y}]^2 $. 
For this dataset, intensity features rank higher than timing features (Figure~\ref{fig:cone}b-c), consistent with previous work~\cite{Atema1996} and predictions are more accurate in the crosswind than in the downwind direction (compare with Figure~S2). For reference, a random guess with flat probability within the correct lower and upper bounds yields $\chi_{\text{random}}=2$; whereas a target function $f_{\text{trivial}}(x)=\langle y\rangle_{\text{test}}$ that learns the average of the output over the test set yields $\chi_{\text{trivial}}=1$.\\
\begin{figure}[h!!]
\begin{center}
\includegraphics[width=0.49\textwidth]{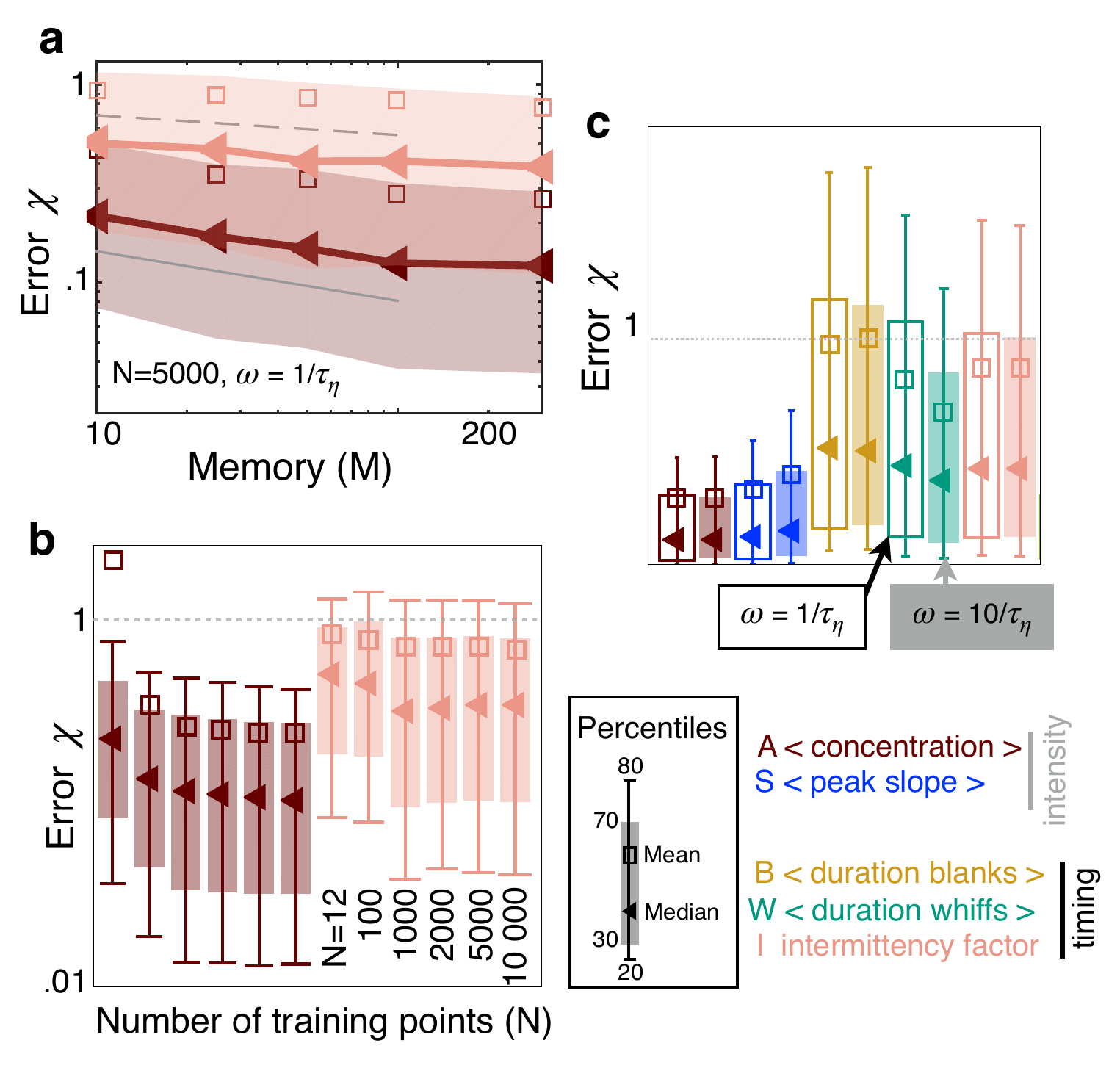}
\caption{The sampling strategy affects performance but not ranking. (a) Error $\chi$ as a function of memory in units of Kolmogorov times $\tau_\eta$; memory is defined as the duration of the time series of odor concentration $\mathbf{c}_i=(c(\mathbf{z}_i,t_i),...,c(\mathbf{z}_i,t_{i+M}))$ used to compute the five features $x_i^1,...,x_i^5$, i.e.~memory$=t_{i+M}-t_i = M/\omega$. Red and pink: Performance using ${\mathbf{x}}_i=x_i^1$ (average concentration) and ${\mathbf{x}}_i=x_i^5$ (intermittency factor). The number of training points and the frequency of sampling are fixed, $N=5000$ and $\omega = 1/\tau_\eta$. Dotted, dashed and solid grey lines are power laws with exponents $-1/5$, $-1/10$ and $-1/4$ respectively to guide the eye. (b) Error as a function of number of points in the training set $N$, with $N_t=13500$ points in the test set, memory$=100\tau_\eta$ and $\omega = 1/\tau_\eta$. Color code as in (a). (c) Performance using the five individual features as input with $N=5000$,  $N_t=13500$, memory$=100\,\tau_\eta$ sampling odor at frequency $\omega=1/\tau_\eta$ (empty bars) and $\omega=10/\tau_\eta$ (filled bars). Key shows color coding.
}
\label{fig:sampling}
\end{center}
\end{figure}

Next we analyze whether and how the sampling strategy affects performance and ranking of the features. Most results are shown for a memory of $100\tau_\eta\approx 15\,s$. Performance improves with longer memory (Figure~\ref{fig:sampling}a), because this allows to better average out noise and obtain more stable estimates of the features. But improvement follows a slow power law so that waiting for example 20 times longer yields predictions only about twice as precise. On the other hand, waiting as little as $10\tau_\eta \approx 1.5$ seconds still allows to make predictions, albeit less precise. We then verify whether performance may improve with a larger training set. Because we infer distance from an individual (scalar) feature, the problem is one dimensional and we find that a small number of training points, which we indicate with $N$, is sufficient to reach a plateau in prediction performance (Figure~\ref{fig:sampling}b). We choose $N=5000$ training points, which is also robust to the case with more than one feature (Supplementary Figure~S3). Finally, sampling more frequently than once per Kolmogorov time does not essentially affect the results nor ranking (Figure~\ref{fig:sampling}c). Similar results hold for the crosswind direction (Supplementary Figure~S4).

\begin{figure}[h!!]
\begin{center}
\includegraphics[width=0.5\textwidth]{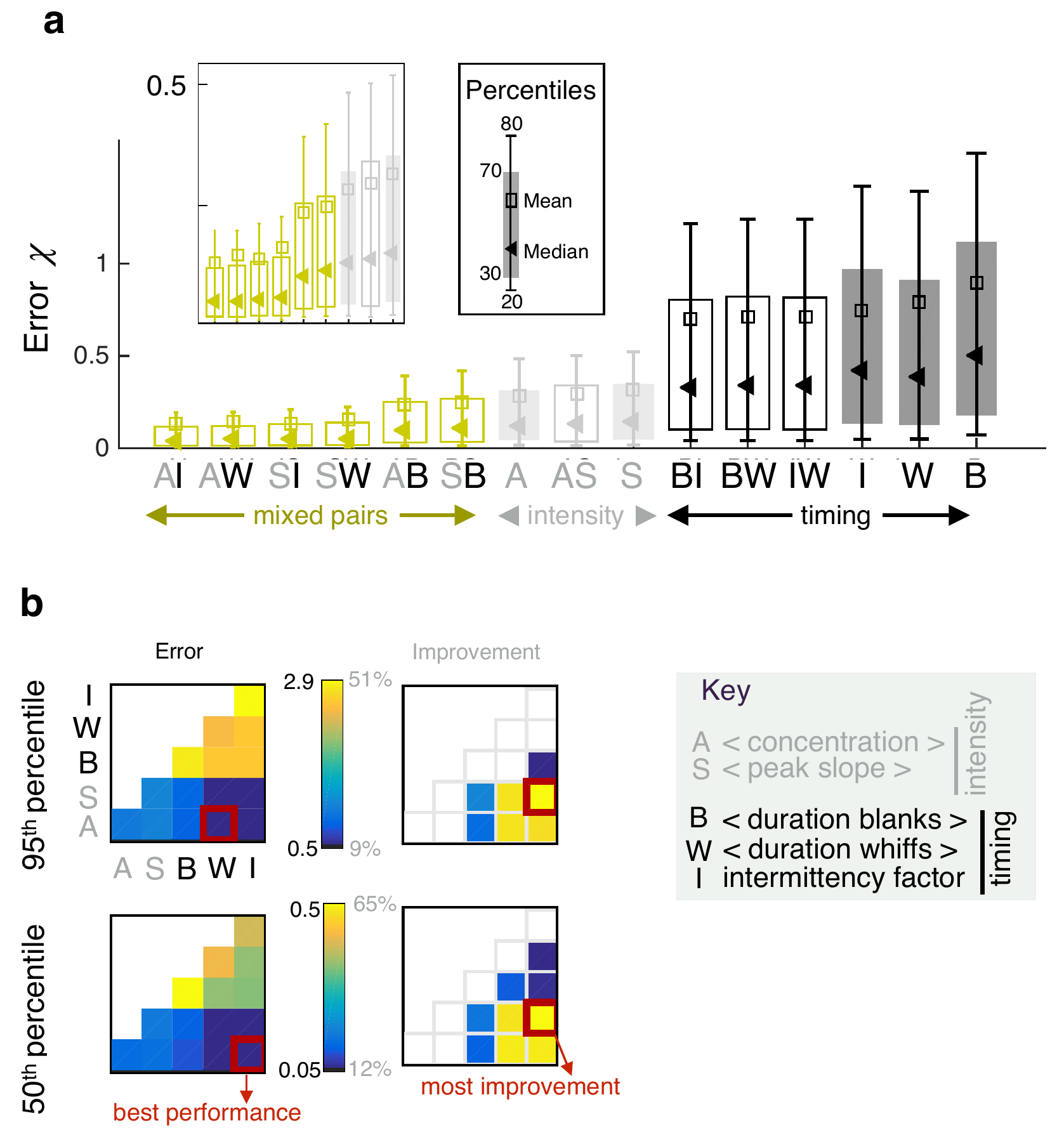}
\caption{Pairing one timing feature and one intensity feature considerably improves performance. 
a) Error $\chi$ obtained with individual features (full bars) and pairs of features (empty bars). Grey and black indicate pairings of two intensity features  and two timing features respectively; green indicates mixed pairs of one timing and one intensity feature. (b) Performance (left) and relative improvement over the best of the two paired features (right). Results for the median (bottom) and the 95$^{\text{th}}$ percentile (top). Within each table plot, rows from bottom to top and columns from left to right are labeled by the 5 individual features: A (average, $x^1$), S (slope $x^2$), B (blanks $x^3$), W (whiffs $x^4$), I (intermittency $x^5$). Results with individual features are shown on the diagonal; results pairing feature $i$ and feature $j$ are shown at position $(i,j)$. Mixed pairs provide both the best performance and the largest improvement over individual features.
}
\label{fig:twofeatures}
\end{center}
\end{figure}
Pairing two observables improves performance in some cases, but not always. In fact, pairing two features of the same category results in little to no improvement (Figure~\ref{fig:twofeatures} and similarly for the crosswind direction, Figure~S5). In contrast, combining one intensity and one timing feature improves performance considerably, up to 65\%. This result can be understood by mapping the error done by individual features in space (Supplementary Figure~S6), showing that intensity and timing features are complementary, i.e.~intensity features perform well in locations where timing features perform poorly. \\

\begin{figure}[h!!]
\begin{center}
\includegraphics[width=0.5\textwidth]{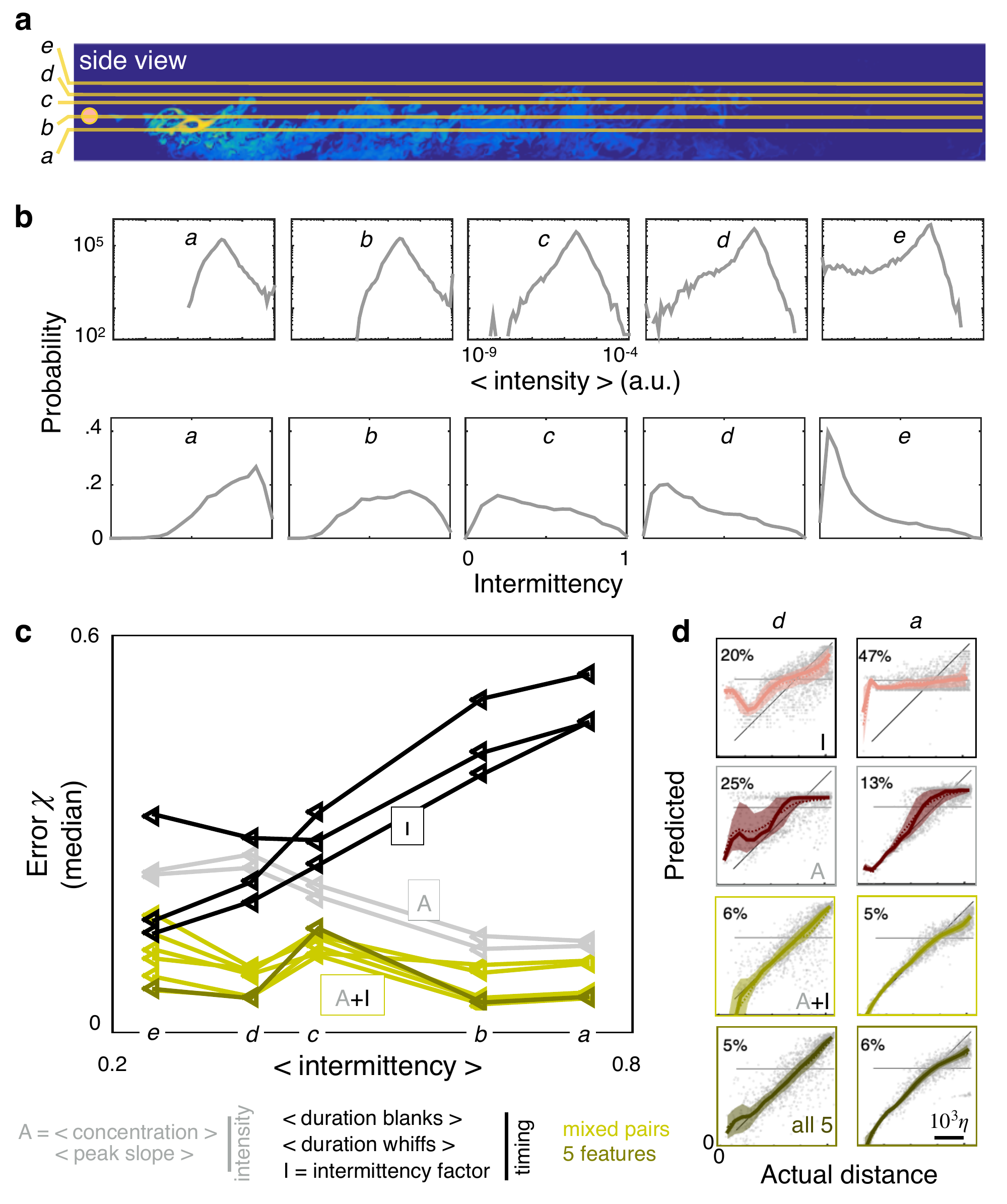}
\caption{Ranking shifts with height from the ground. (a) Datasets \emph{a} to \emph{e} correspond to data obtained at heights $z/H=25\%$, $37.5\%$, $50\%$, $55\%$ and $65\%$ respectively. (b) Distribution of intensities (top) and intermittency factors (bottom) over the training set from \emph{a} to \emph{e} (left to right). Moving away from the boundary, the odor becomes less intense and more sparse. (c) Median performance as a function of average intermittency factor of the training set for individual intensity (grey) and timing (black) features, mixed pairs of one intensity and one timing feature (green) and all five features together (dark green). (d) Predicted \emph{vs} actual distance, to visualize a representative subset of the results in (c), scale bar $10^3\eta$. Ranking depends sensibly on height: intensity features outperform timing features near the substrate, where there is more odor and it is more continuous; timing features outperform intensity features further from the substrate where there is less odor and it is more sparse; mixed pairs perform best across all conditions; combining five features provides little to no improvement over mixed pairs. }
\label{fig:z}
\end{center}
\end{figure}

We next seek to clarify whether the results depend on space. To this end we compose five different dataset, \emph{a} to \emph{e}, obtained by extracting odor snapshots from horizontal planes at source height (\emph{b}), above the source (\emph{c} to \emph{e}), and below the source (\emph{a}) (Figure~\ref{fig:z}a). From \emph{a} to \emph{e}, sparsity increases and intensity decreases (Figure~\ref{fig:z}b) simply because closer to the boundary, the air slows down and the odor accumulates. By analyzing performance across these dataset, we find that ranking of individual features shifts considerably. The two intensity features outperform all timing features when the dataset is not very sparse (dataset \emph{a}-\emph{b}, Figure~\ref{fig:z}c, d left). In contrast, two timing features (intermittency factor and blank duration) outperform all others for the more sparse and less intense dataset \emph{d-e} (Figure~\ref{fig:z}c, d right). Whiff duration performs poorly in \emph{d-e} because intermittency is too severe and whiffs are short in duration thus bare little information (the average whiff duration is 1 to 7 time steps in over 90\% of the time series). Although the ranking of individual features shifts with height, pairing one intensity and one timing feature remains the most successful strategy across all heights (Figure~\ref{fig:z}c,d). In contrast, combining all five features contributes little improvement (Figure~\ref{fig:z}c-d and Supplementary Figure~S7).

\begin{figure}[h!!]
\begin{center}
\includegraphics[width=0.5\textwidth]{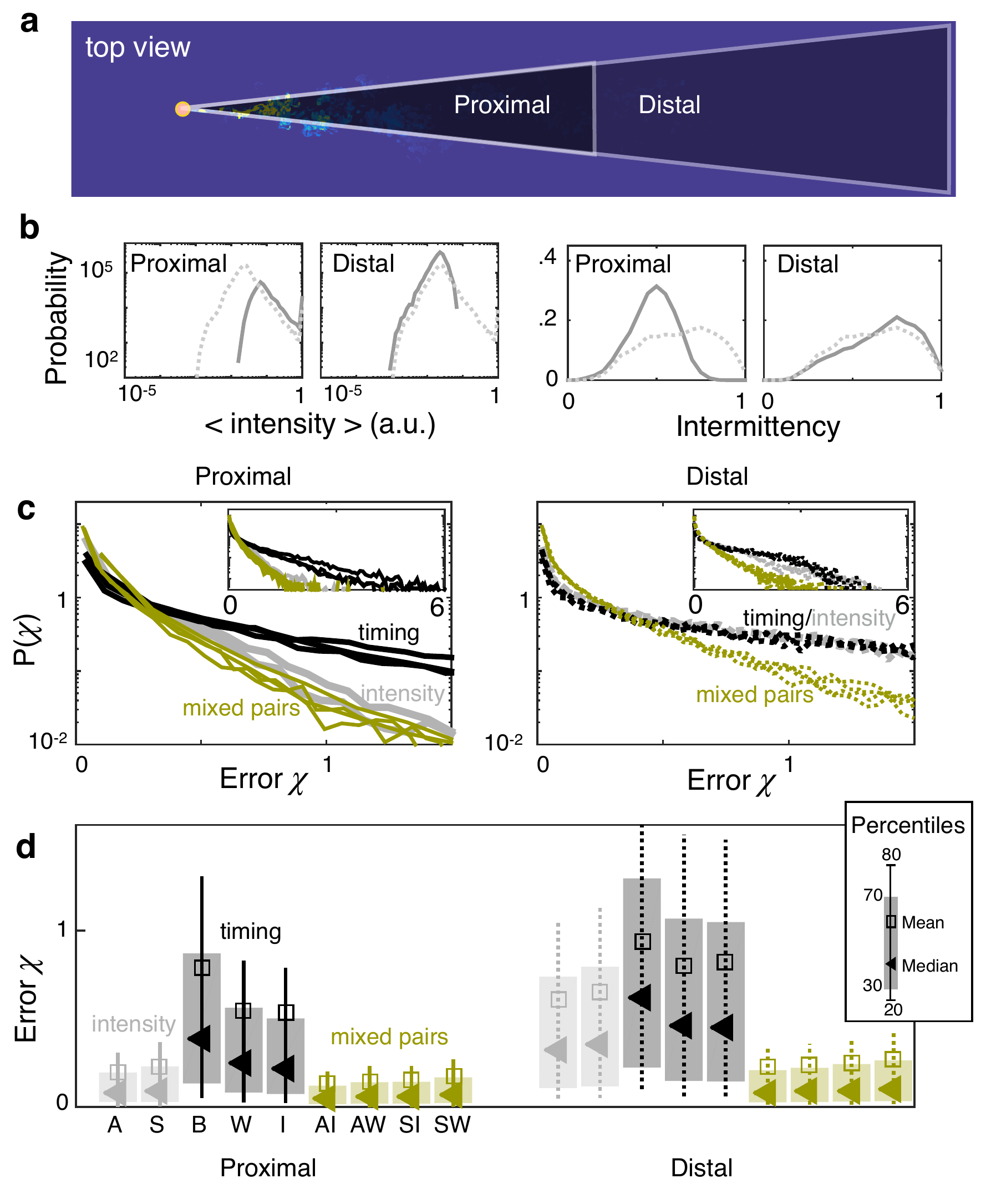}
\caption{Ranking depends on distance from the source. (a) At source height, the dataset is split in proximal (distance<2330$\eta$) and distal (distance>2330$\eta$). (b) Distributions of average odor intensity (left) and intermittency factor (right) over the training set; closer to the source, the odor is more intense and more sparse. (c) Distribution of test error for the proximal (left) and distal (right) problem showing intensity features (grey) outperform timing features (black) at close range, but not in the distal problem where differences in the error distribution are limited to the tails (see insets). Mixed pairs of features (green) outperform individual features either marginally (left) or considerably (right). (d) Percentiles of the error distribution in (c) for the proximal (left) and distal (right) problems confirming the picture emerged from (c).}
\label{fig:cf}
\end{center}
\end{figure}

Let us now focus on the plane at source height 
and separate locations based on their distance from the source. We assemble a distal dataset and a proximal dataset, composed of points that are further and closer than 2330$\eta$ from the source respectively (Figure~\ref{fig:cf}a). The odor is more intense and more sparse closer to the source and it becomes more dilute and less sparse with distance from the source (Figure~\ref{fig:cf}b). Performance of individual features degrades with distance (Figure~\ref{fig:cf}d). Intensity features clearly outperform timing features at close range, as seen both from various percentiles of the test error (Figure~\ref{fig:cf}d, left) as well as the full distribution (Figure~\ref{fig:cf}c, left).
The disparity between timing and intensity features disappears in the distal problem: the error distribution for all individual features is essentially superimposed except for the tails (Figure~\ref{fig:cf}c, right and inset), which cause small differences in the median and other percentiles of the error (Figure~\ref{fig:cf}d, right). Remarkably, mixed pairs outperform all individual features in both the distal and proximal problems (Figure~\ref{fig:cf}c-d).
In the aggregate, results demonstrate that, even within a single turbulent flow, ranking shifts considerably. 
Namely, measuring timing of odor encounters is most useful in regions where the odor is dilute, i.e.~far from the source and from the substrate;  whereas measuring intensity is most useful in concentrated conditions, i.e.~close to the source or the substrate. 

\section*{Discussion}
Our results demonstrate that within the cone of detection, the time course of an odor bares useful information for source localization even at meters from the source. We find that the concentration and the slope of a turbulent odor signal, averaged over a memory lag, are particularly useful to predict source location at close range or near the boundary. These features quantify the intensity of the odor and its variation. The primacy of the intensity features wanes in more challenging conditions, e.g.~moving away from the source or away from the boundary. In these portions of space, where the odor is scarcer, features that quantify timing of odor detection become as effective as intensity features, or more effective. As mentioned above, to compute timing features, it is sufficient to record \emph{when} the odor is on, rather than its intensity, suggesting neural representations that binarize odor as an on-off signal may emerge in organisms that evolved to use olfaction from large distances. Interestingly, nearly binary representations have been found in insects, which are arguably the best studied example of long range olfactory navigation. \\
Note that while the statistics of an odor plume clearly depends on all details of the flow and the source, see e.g.~\cite{Justus2002,prx,fackrell_robins}, here we keep all of these parameters constant and demonstrate that even within a single flow, odor dynamics and the best predictors vary considerably in space.
This begs the next question: do organisms switch between different modalities depending on attributes of odor dynamics, which will vary in space? This could be the case for mice, where the neural activity in the first relay of olfactory processing does in fact depend on how sparse is the odor~\cite{Lewisetal2021}. Specifically, sparse odor cues elicit individual responses that follow closely the ups and downs of the odor in time. In contrast, continuous signals elicit intense responses which are however uncorrelated to the temporal dynamics of the odor itself~\cite{Lewisetal2021}. \\

We find that features within the same class are redundant whereas features from different classes are complementary. Indeed, features of the same class have similar patterns of performance in space, but each class has a distinct pattern. 
As a consequence, measuring both timing and intensity is beneficial, but using more than one feature to quantify either timing or intensity provides no advantage. Combining all features does not improve over the performance of mixed pairs, consistent with redundancy within each class. Note that there is no fundamental reason to expect features from the same class to be redundant, and further work with a larger library of features is needed to prove or disprove this notion.\\ 

Importantly, mixed time/intensity pairs of features outrank individual features robustly, i.e.~in all portions of space, regardless of distance from the source and from the ground. This is in contrast with individual features and suggests relying on simultaneous timing and intensity features is advantageous when odors are sensed at various distances from the source and from the substrate. Interestingly, the coexistence of bursting olfactory neurons and canonical  
olfactory neurons in lobsters suggests these animals are in fact able to measure simultaneously timing and intensity \cite{Parketal2014,Acheetal2016}, which is consistent with the increased predictive power of the mixed pairs of features. 
Similarly, in mammals, optogenetic activation of the olfactory bulb~\cite{smearetal2013} demonstrates that both kinds of measures guide behavior (lick \emph{vs} no lick). \\

In this work, we have investigated the problem of predicting the location of a target from measures of the time course of a turbulent odor. 
Previous work explored a related question, i.e.~how to best represent instantaneous snapshots of the odor to encode maximum information about source location~\cite{victoretal2019}. 
The two approaches are not immediately comparable: first, \cite{victoretal2019} consider few snapshots of the odor, rather than measures of its time course. Second, maximizing information does not guarantee good predictions (to make predictions information needs to be extracted and processed, and importantly the focus is on new data that were not previously seen). We provide two comments that are relevant if information is the limiting factor for prediction accuracy: \emph{(i)} binary representations were suboptimal in all conditions considered in~\cite{victoretal2019,boieetal2018}, i.e.~at few tens of cm from the source. This is consistent with our results in concentrated conditions, where timing features -accessible through binary representations- are suboptimal. Our evidences suggest however that the result may not hold in more dilute conditions, where the gap between binary and more accurate representations should become increasingly small. 
\emph{(ii)} Individual snapshots of odor from~\cite{victoretal2019,boieetal2018} contained 1 to 2 bits of information about source location, but allocating more resources to represent how the odor varies in time was found informative~\cite{victoretal2019,boieetal2018}. Our mixed pairs of features at close range achieve precisions of 5\% to 6\%, corresponding to coding for position with words of 4 to 4.3 bits. Our results thus confirm that memory is indeed useful, but the gain does not increase indefinitely with further memory.
\\

The literature on olfactory navigation is vast. Although a complete review of available algorithms is beyond the scope of the present work, we remark that recent results investigated gradient descent algorithms using either concentration alone~\cite{Gireetal2016}, or various measures of timing and intensity~\cite{Parketal2016,michaelisetal2020,Leathers2020}. Overall, both intensity and timing appear to have a potential to lead to an odor source, consistent with our results on individual features. A combination of the two kinds of features was found beneficial in~\cite{Leathers2020}, consistent with our results on mixed pairs. Whether good predictors may be good variables for navigation in more general contexts remains to be understood. 
\\



Here we have analyzed the features that enable the most accurate prediction of source location. We add a few observations about the significance of the results for animal behavior. First: whether animals rely on features from either class will depend on what features best support behavior. It is often implicitly assumed that features that bare reliable information on source location are also the most useful for navigation. However, this connection between prediction and navigation is far from straightforward and more work is needed to establish whether accurate predictions imply efficient navigation. Second: animals are unlikely to have prior information on the details of the odor source, e.g.~its intensity. Timing features are more robust than intensity features with respect to the intensity of the source and may thus be favored regardless of their performance, which was argued in~\cite{Schmukeretal2016}. In our work, timing features are precisely invariant with source intensity because we define the detection threshold adaptively (see Materials and Methods). 
More realistic conditions will need to be evaluated, where dependence on source intensity emerges as a result of non-linearities that we did not model in this work. These effects emerge for example, close to a boundary which partially absorbs the odor~\cite{emonet_controlling_stimuli}, or in the case of fixed thresholds, although this dependence is weak in the far field where timing features are most useful~\cite{prx}. 
Third: we have focused on predicting source location from within the cone of detection, where an agent will detect the odor quite often. However, a crucial difficulty of turbulent navigation is to find the cone itself. We cannot address the problem of predicting source location from outside the cone because detections are so rare that we lack statistics. The distinction between inside and outside the cone of detection is key for navigation with sparse cues~\cite{Reddy2021} and deserves further attention.


\matmethods{

\vspace*{-0.5cm}\subsection*{Direct numerical simulations of turbulent odor plumes}
\begin{table*} [h!]
\center
\caption{\footnotesize 
Parameters of the simulation. Length $L$, width $W$, height $H$ of the computational domain; horizontal speed along the centerline $U$;  mean horizontal speed $U_b=\langle u \rangle$; Kolmogorov length scale $\eta=(\nu^3/\epsilon)^{1/4}$ where $\nu$ is the kinematic viscosity and $\epsilon$ is the energy dissipation rate; mean size of gridcell $\Delta x$; Kolmogorov timescale $\tau_\eta=\eta^2/\nu$; energy dissipation rate $\epsilon= \nu/2 \langle(\partial u_i/\partial x_j +\partial u_j/\partial x_i)^2\rangle$; 
Reynolds number $Re=U (H/2) / \nu$ based on the centerline speed $U$  and half height; Reynolds number $Re_\lambda = U \lambda/\nu$ based on the centerline speed and the Taylor microscale $\lambda$; magnitude of velocity fluctuations $u'$ relative to the centerline speed; large eddy turnover time $T = H/2u'$.
First row reports results in non dimensional units; second and third rows correspond to dimensional parameters in air and water assuming the velocity of the centerline is 50 cm/s in air and 12 cm/s in water.}
\label{tab:par}
\setlength\tabcolsep{3pt}
\begin{tabular}{|c c c c c c c c c c c c c c c c c|}
\toprule
& $L$ & $W$ & $H$ & $U$ &  $U_b$ & $\eta$ & $\Delta x $ & $ \tau_\eta$ &  $\epsilon$ & $\lambda$ & $y^+$ & Re & Re$_\lambda$ & $u'/U$ & $T$  \\
\midrule
\midrule
          & 40 & 8 & 4 & 32  & 23 & 0.006 & 0.025 & 0.01  & 39 & 0.17 &  0.0035 & 16000 & 1360 & 11\%&$64\tau_\eta$  \\
\midrule
\midrule
air      & 9.50 m & 1.90 m & 0.96 m & 50 cm/s   & 36 cm/s&  0.15 cm& 0.6 cm& 0.15 s & 6.3e-4 m$^2$/s$^3$& 4 cm& 0.09 cm& &  \\
water & 2.66 m & 0.53 m & 0.27 m & 12 cm/s  & 8.6 cm/s&  0.04 cm& 0.2 cm& 0.18 s & 3e-5 m$^2$/s$^3$ & 1 cm&0.02 cm & & \\
\bottomrule
\end{tabular}
\end{table*}

To reproduce a realistic odor landscape and generate the dataset showed in Figure~\ref{fig:snapshots}, we solve the Navier-Stokes~\eqref{eq:ns} and the advection-diffusion equation for passive odor transport~\eqref{eq:ps} at all relevant scales of motion from the smallest turbulent eddies (Kolmogorov scale $\eta$) to the integral scale ($L > 600\eta$), using Direct numerical simulations (DNS):
\begin{equation}{
    \partial_t \mathbf{u} + \mathbf{u} \cdot \nabla \mathbf{u} = -\frac{1}{\rho}\nabla P +\nu \nabla^2 \mathbf{u}  \quad \quad \nabla \cdot \mathbf{u} = 0
    }
    \label{eq:ns}
\end{equation}
\begin{equation}{
    \partial_t \theta + \mathbf{u} \cdot \nabla \theta =\kappa_\theta \nabla^2 \theta + q
    }
    \label{eq:ps}
\end{equation}
\noindent where $\mathbf{u}$ is the velocity field, $\rho$ is the fluid density, $P$ is pressure, $\nu$ is the fluid kinematic viscosity, $\theta$ is the odor concentration, $\kappa_\theta$ is its diffusivity and $q$ an odor source.  
We simulate a turbulent channel flow with a concentrated odor source and an obstacle that generates turbulence by customizing the open-source software Nek5000 \cite{Nek5000} developed at Argonne National Laboratory, Illinois. 
Nek5000 employs a spectral element method (SEM) \cite{Patera1984} \cite{Orszag1980} based on Legendre polynomials for discretization \cite{Ho1989}, and a 4th order Runge-Kutta scheme for time marching. The code is written in fortran77 and C and it uses MPI for parallelization. 
 
The three dimensional channel is divided in $E = 160\,000$ discrete elements: $200 \times 40 \times 20$ (number of elements in length $\times$ width $\times$ height); within each element the solution is expanded in 8th grade tensor-product polynomials so that the domain is effectively discretized in 81\,920\,000 elements. The average spatial resolution is equal in each direction ${\Delta x} \approx 4\eta$. A cylindrical cap of height~$= 160 \eta$ is added on the ground; the cylinder spans the entire width of the channel. 
The mesh is adapted to fit the cylinder. Fluid flows from left to right and the obstacle generates turbulence in the channel, in particular the height of the cylinder tunes the velocity fluctuations. The velocity fluctuations are defined as $\delta u(\mathbf{z},t)=u(\mathbf{z},t)-\langle u(\mathbf{z},t) \rangle_y$; their intensity is $u'=\sqrt{\langle (\delta u)^2 \rangle}$, where averages are intended in space and time. Table~\ref{tab:par} summarizes the parameters that characterize turbulence. 
 
Each simulation runs for 300 000 time steps where $\delta t = 10^{-2}\tau_\eta$ and follows from a severe Courant criterium with $U \Delta t/\Delta x < 0.4$ to ensure convergence of both the velocity and scalar fields. Snapshots of velocity and odor fields are saved at constant frequency $\omega = 1/\tau_\eta$ (except for results in Figure~\ref{fig:sampling}c where snapshots are saved 10 times more frequently). Each DNS requires 2 weeks of computational time using 320 cpus. \\

\subsection*{Boundary conditions and odor source} 
We impose a Poiseuille velocity profile at the inlet: $\mathbf{u}=(u,0,0)$ and $u = -(z_3^2 - z_3)/16\nu$, where $z_3$ is the vertical coordinate. We set a no-slip condition $\mathbf{u}=0$ at the ground and on the obstacle; on the remaining boundaries we impose the turbulent outflow condition defined in~\cite{FISCHER2007} that imposes a positive exit velocity to avoid potential negative flux and the consequent instability it generates. 
More precisely, the divergence ramps up from zero to a positive value along the element closest to the boundary: $\nabla\cdot \mathbf{u} = C[1-(z_\perp/ \Delta x)^2]$, where $z_\perp$ is the distance from the boundary and $C=2$ is the minimal value that ensures convergence. 
For the odor, we impose a Dirichlet condition ($\theta = 0$) at the ground, on the obstacle and at the inlet; while an outflow condition is set at the top, on the sides and at the outlet: $k(\nabla T)\cdot \mathbf{n}=0$. We introduce a source located right above and downstream of the obstacle, at coordinates $x_s$ = 810$\eta$, $y_s$ = 650$\eta$, $z_s$ = 238$\eta$; odor intensity at the source is defined by a gaussian distribution $q = e^{[(z_1-x_s)^2+(z_2-y_s)^2+(z_3-z_s)^2]/(2\sigma ^2)}$, where $\sigma = 5\eta$. \\

\subsection*{Machine learning}
To learn the correct position of a target source given an odor,   we propose to use supervised  machine learning. We next review some key ideas, and refer to standard textbooks  for further details e.g.~\cite{hastietibshirianifridman}.

The goal in supervised  learning is  to infer a function $f$ given  a training set $(\mathbf{x}_1,y_1), \dots (\mathbf{x}_N, y_N) $ of input/output pairs. A good function estimate should allow to {\em predict} the outputs associated to  {\em new}  input points. In our setting each input $\mathbf{x}$ is a 1-, 2- or 5- dimensional vector whose entries are scalar features of odor time series, where the odor is sampled at a specific spatial location. From every sampling location, we compute the distance to the source and this distance is the output $y$.

To  measure  how close the prediction $f(\mathbf{x})$ is  to the correct  output $y$, we consider  the square loss  $(f(\mathbf{x})- y)^2$. Following a statistical learning framework, the data are assumed to be sampled according to a fixed but unknown data distribution $P$. In this view,  the ideal solution   $f^*$ should  minimize the expected loss $\langle l(f(\mathbf{x}),y) \rangle$ over all data distributed according to $P$. In practice, only an empirical loss based on training data can be measured, and the search for a solution needs be restricted to a suitable class of hypothesis. Note that,  the choice of the latter is critical since the nature of the function to be learnt is not known a priori. A basic choice is considering linear functions $f(\mathbf{x})=\mathbf{w}\cdot\mathbf{x}$.
In this case, minimizing the training loss reduces to linear least squares $\text{min} \frac{1}{N}||Y- X\cdot\mathbf{w}||^2$, where $X$ is the matrix composed of the $N$ training data input $X=(\mathbf{x}_1,...,\mathbf{x}_N)^T$ and $Y$ is the vector composed of the $N$ labels of the training set $Y=(y_1,...,y_N)^T$.  The corresponding solution is easily shown to be  $\mathbf{w}=(X^TX)^{-1} X^T Y$. In Figure S1, we  show that the choice of linear models has limited predictive power and does not allow to rank features. To tackle this issue we consider kernel methods \cite{smola}, a more powerful  class of nonlinear models corresponding to functions of the form $f(\mathbf{x}) =\sum_{i=1}^N k(\mathbf x_i, \mathbf x) {\mathbf c}_i.$
Here, $k(\mathbf x , \mathbf x ')$ is a so called kernel, that here we will choose to be the Gaussian kernel $k(\mathbf{x},\mathbf{x}') = e^{-{\parallel \mathbf{x}-\mathbf{x}'  \parallel^2}/{2\sigma^2}}$. The coefficients ${\mathbf c}=({\mathbf c}_1,\dots,{\mathbf c}_n)$  are given by the expression 
\begin{equation}
\label{eq:solut}
{\mathbf c}= (K+\lambda N I  )^{-1}{\mathbf y}
\end{equation}
which minimizes
$$
\frac 1 N \|K {\mathbf c} - {\mathbf y}\| +\lambda {\mathbf c}^\top K {\mathbf c}. 
$$
In the above expression $K$ is the $N$ by $N$ matrix with entries $K_{ij}=  k ({\mathbf x}_i, {\mathbf x}_j)$. The first term can be shown to be a data fit term whereas the second term can 
be shown to control the regularity of the obtained solution \cite{smola}.  The {\em regularization parameter} $\lambda$ balances out the two terms and needs be tuned, together with the kernel parameters (the Gaussian width $\sigma$ in our case). \\
Kernel methods offer a number of advantages. They are nonlinear, and hence can learn a wide range of complex input/output behavior. They are an example of nonparametric models, where the complexity of the model can adapt to  the problem at hand and indeed learn any kind of continuous function provided enough data. This can be contrasted to linear models that clearly cannot learn any nonlinear function. Moreover, by tuning the hyper-parameters $\lambda, \sigma$ more or less complex shape can be selected. When $\lambda$ is small we are simply fitting the data, possibly at the price of stability, whereas for large $\lambda$ we are favoring simpler models. With small $\sigma$ we allow highly varying functions, whereas with large enough $\sigma$ we essentially recover linear models. 

Indeed, the choice of these parameters is crucial and  tested and visualized in Figure~S8. Here it is shown that for $\lambda\rightarrow 0$, the solution incurs in the well known stability issues for large $\sigma$ and overfitting issues for small $\sigma$. 
  We note  that ideally one would want to choose these hyper-parameters minimizing the test error, however this would lead to overoptimistic estimates of the prediction properties of the obtained model. Hence, we consider a hold-out cross validation protocol, where the training data are further split in a training and a  validation sets. The new training set is used to compute solutions corresponding to different hyper-parameters. The validation set is used as a proxy for the text error  to select the hyper-parameters with small corresponding error. The prediction properties of the model thus  tuned is then assessed on the test set. 
%

\subsection*{Dataset} 
To compose the dataset for regression we first extract two-dimensional snapshots of odor at fixed height from the 3D simulation. Each snapshot from the simulation has dimensions $1600\times320$ (number of points in the downwind direction $\times$ crosswind direction). The initial evolution up to $300\,\tau_\eta$ is excluded from the analysis as odor has not yet reached a stationary state. At stationary state we save 2700 frames at frequency $\omega=1/\tau_\eta$ per simulation.  Thus at each spatial location we have the entire time evolution composed of 2700 time points at regular intervals of $\tau_\eta$. We partition each simulation in fragments with $M$ snapshots (duration $M\tau_\eta$). Most simulations are shown for $M=100$, thus for each spatial location we have 27 time series of the same duration (except for results leading to Figure~\ref{fig:sampling}a, where we vary memory from $10\tau_\eta$ to $250\tau_\eta$ resulting in 270 to 10 time series per location respectively). 

The characteristic shape of the odor plume is a cone (Figure~\ref{fig:cone}), that we defined as the region where the probability of detection computed over the entire simulation is larger than 0.35. The training set and test set are obtained by extracting $N=5000$ (unless otherwise stated) and $N_{\text{t}}=13500$ time series portions of duration $M\tau_\eta$. To select these $M$-long time series we extract random locations $\mathbf{z}_i$ to cover homogeneously the cone, i.e.~with flat probability within the cone, and random initial times $t_i$, with the training in the first half of the time history and the test in the second half of the time history. 
Time series that remain entirely under threshold are excluded. 

Each odor time series is further processed by computing five features, two of which quantify intensity of the odor and rely on a precise representation of odor concentration (average concentration and average peak slope) and three of which quantify timing of odor encounters and are computed after binarizing the odor (average whiff and blank duration and intermittency factor). 
The threshold $c_{thr}$ used for binarization is adaptive i.e. $c_{thr}=0.5\langle c|c>0 \rangle$, where the average is computed over each time series separately. The threshold thus varies from $c_{thr} = 0.5 c_0$ at the source to $c_{thr} = 10^{-6}c_0$ at the farthest edges of the cone, where $c_0$ is the concentration at the source. The choice of an adaptive threshold was suggested in~\cite{Gorur-Shandilya2017}.
The precise value of the relative threshold has little effect on the results as shown in Figure~S9, left. Fixed thresholds were tested and discarded because results depend sensibly on the threshold and the optimal threshold varies with the dataset in non-trivial ways (Figure~S9, right). Finally, adaptive thresholds that are defined based on purely local information appears more plausible for a biological system that has no information on the intensity of the source. 

The parameters $\lambda$ and $\sigma$ are obtained through 4-folds cross validation: 
the training set is split in 4 equal parts, 3 are used for training and 1 for validation. The empirical risk is computed on the validation set and averaged over the 4 possible permutations, systematically varying the hyperparameters $\lambda, \sigma$. The couple of hyperparameters that minimize the empirical risk over the validation set is selected through grid search using an $8\times 8$ regular grid and further refined with a $4\times 4$ subgrid. Results are insensitive to further refinement because there is a large plateau around the minimum, as shown in Figure~S2. The optimal hyperparameters are used to compute the solution~\eqref{eq:solut}. The error $\chi$ used throughout the manuscript is simply the normalized test error $\chi =  \sum_{i=1}^{N_t}[y_i-f(\mathbf{x}_i)]^2 /  \sum_{i=1}^{N_t}[y_i - \bar{y}]^2 $. For most of the figures, we used 5000 training points which allows to directly invert the data matrix to obtain the solution~\eqref{eq:solut}. When testing the effect of $N$, we implemented Kernel ridge regression using FALKON\cite{FALKON}, a fast algorithm for matrix inversion (the number of iterations is set to 5 and the number of Nystrom centers is equal to the number of points in the training set).
}
\showmatmethods{} 
\acknow{
This work was supported by the Air Force Office of Scientific Research under award number FA8655-20-1-7028; 
by the French government, through the UCA$^{\textrm{\sc \sf \tiny JEDI}}$ Investments in the Future project managed by the National Research Agency (ANR) under reference number \#ANR-15-IDEX-01; by the project DynCellPol (ANR-19-CE13-0004); by CNRS PICS ``2FORECAST'' and by the Thomas Jefferson Fund a program of FACE.  The authors are grateful to the OPAL infrastructure from Universit\'e C\^ote d'Azur and the Universit\'e C\^ote d'Azur's Center for High-Performance Computing for providing resources and support. N.M.~ and N.R.~are thankful for the support of Instituto Nazionale di Fisica Nucleare (INFN) Scientific Initiative SFT: Statistical Field Theory, Low-Dimensional Systems, Integrable Models and Applications.
}

\showacknow{} 

\bibliography{refs}

\end{document}